\documentclass[11pt]{article} 

\usepackage[utf8]{inputenc} 

\usepackage{geometry} 
\usepackage{graphicx} 

\usepackage{booktabs} 
\usepackage{array} 
\usepackage{paralist} 
\usepackage{verbatim} 
\usepackage{subfig} 
\usepackage{amsmath} 
\usepackage{amssymb} 
\usepackage{url} 
\usepackage{epstopdf} 
\usepackage{color}
\usepackage{stackengine}
\usepackage{graphicx}
\usepackage{cite}

\usepackage{fancyhdr} 
\usepackage{sectsty}
\allsectionsfont{\sffamily\mdseries\upshape} 

\usepackage[nottoc,notlof,notlot]{tocbibind} 
\usepackage[titles,subfigure]{tocloft} 

\title{Limits of Predictability of Cascading Overload Failures in Spatially-Embedded Networks with Distributed Flows}
\author{A. Moussawi$^{1,2}$,  N. Derzsy$^{1,2}$, X. Lin$^{2,3}$, B. K. Szymanski$^{2,3}$, G. Korniss$^{1,2}\footnote{E-mail: korniss@rpi.edu}$}

\begin{document}
\maketitle

\begin{flushleft}
$^{\bf{1}}$ Department of Physics, Applied Physics, and Astronomy, Rensselaer Polytechnic Institute,
110 8$^{th}$ Street, Troy, NY, 12180-3590 USA \\
$^{\bf{2}}$ Social Cognitive Networks Academic Research Center,
Rensselaer Polytechnic Institute, 110 8$^{th}$ Street, Troy, NY, 12180-3590 USA \\
$^{\bf{3}}$ Department of Computer Science,
Rensselaer Polytechnic Institute, 110 8$^{th}$ Street, Troy, NY, 12180-3590 USA \\
\end{flushleft}


\section*{Abstract}

Cascading failures are a critical vulnerability of complex information or infrastructure networks. Here we investigate the properties of load-based cascading failures in real and synthetic spatially-embedded network structures, and propose mitigation strategies to reduce the severity of damages caused by such failures. We introduce a stochastic method for optimal heterogeneous distribution of resources (node capacities) subject to a fixed total cost. Additionally, we design and compare the performance of networks with \textit{N}-stable and \textit{(N-1)}-stable network-capacity allocations by triggering cascades using various real-world node-attack and node-failure scenarios. We show that failure mitigation through increased node protection can be effectively achieved against single node failures. However, mitigating against multiple node failures is much more difficult due to the combinatorial increase in possible failures. We analyze the robustness of the system with increasing protection, and find that a critical tolerance exists at which the system undergoes a phase transition, and above which the network almost completely survives an attack. Moreover, we show that cascade-size distributions measured in this region exhibit a power-law decay. Finally, we find a strong correlation between cascade sizes induced by individual nodes and sets of nodes. We also show that network topology alone is a weak factor in determining the progression of cascading failures.

\section*{Introduction}

In complex information or infrastructure networks even small localized disruptions can give rise to congestion that can trigger cascading failures (avalanches), a sequence of consecutive failures \cite{Bernstein_2011}. For example in transportation systems, when a component fails the transportation flow is redistributed on the remaining network, which can cause subsequent congestion and failures. This phenomena can be observed in various flow-driven systems such as infrastructure networks \cite{blackout_2003,blackout_2011}, urban microgrids \cite{Gonzalez_2016}, commuter transportation and mobility networks \cite{Barabasi_2012,Neda_2013,Toroczkai_2014}, financial systems \cite{Sachs_2009}, and biological networks \cite{Anrather_2011}. Among infrastructure networks a great interest is focused on the study of cascading failures occurring in electrical power grids \cite{Hines_2006,Hines_,Hines_2009,Soltan_2014,Zussman_2016,Verma,Dobson,Rahnamay}. Due to the latest technological advances, our modern society permanently relies on continuous and stable electricity provided by power grid systems. Any localized disruption can induce a cascading failure in the power grid, resulting in blackouts often over extended geographical regions, which can create life-threatening situations, and also cause substantial financial losses. Cascading failures propagate at very high speeds, rendering real-time mitigations impractical in the event of a disruption.

Prior studies on cascading failures in power grids suggest that power grid blackouts follow a first-order phase transition \cite{Scala}. Similarly, a recent study conducted on the US Western Interconnect power grid has revealed that the probability of cascading failures occurring, with respect to the value of the tolerance parameter, exhibits a bimodal nature, characteristic to first-order transitions \cite{Buldyrev_2016}. These results indicate that a disruption in the system can either cause insignificant damage or trigger severe cascading failures. Another study conducted on scale-free networks has shown that within this phase transition region the number of failed nodes exhibits a power-law tail distribution \cite{Lee}. Other studies have also reported a power-law behavior in the distribution of blackouts in power grids, phenomena related to self-organized criticality \cite{Carreras_2002}. Similarly, the size distribution of cascading failures in the North American power grid \cite{Hines_2009}, as well as the size distribution of power blackouts for the Norwegian and North American power grids \cite{Kertesz_2006} also follow a power-law. Moreover, the power-law behavior has been detected in neuronal networks processing information, such as in the size distribution of neuronal avalanches \cite{Klaus_2011}.

In a more generalized context, power grid systems are spatially-constrained networks, characterized by short-range links. Asztalos, et al.  \cite{Asztalos_2014} have shown that an extensively studied standard cascade mitigation strategy of preemptively disabling a fraction of underperforming nodes \cite{Motter_2004} is largely ineffective in spatially constrained networks. In addition, their study on the European power grid network \cite{Asztalos_2014} has revealed a lack of self-averaging in cascade processes, and showed that increasing the excess capacity does not reduce the severity of the cascade in a monotonic fashion. Further analysis suggested that conventional mitigation strategies are largely ineffective. Therefore, preventive measures, such as designing a stable system against failures or identifying and increasing protection in more vulnerable regions, are crucial to minimize the probability of system disruption.

Motivated by these prior results, we study a model of load-based cascading failures on spatial networks. In particular, we use the European power transmission network maintained and operated by the Union for the Co-ordination of Transmission of Electricity (UCTE), and study it as a random resistor network carrying distributed flow \cite{Lopez_2005,Korniss_2006}. We also employ Random Geometric Graphs (RGGs) \cite{Penrose_2003,Dall_2002} for finite-size scaling analysis. Prior studies focused on resistor networks studied flow optimization, transport efficiency, and vulnerability in complex networks \cite{Korniss_2006,Korniss,Asztalos_2012,Buldyrev_2016}.
We analyze the behavior of cascading failures in the UCTE network, and propose mitigation strategies to reduce the vulnerability of the system against random failures or targeted attacks. We find that mitigation against single node failure can be effectively achieved, however as the complexity of disruption (number of failed nodes) increases, the predictability of the resulting damage becomes increasingly curtailed.

\section*{Results}

Our current work is centered on the study of spatial networks, namely the Union for the Co-ordination of Transmission of Electricity (UCTE) data set \cite{UCTE_data} representing the power grid system of continental Europe in year 2002. The network comprises $N=1254$ transmission stations and $E=1812$ edges spanning 18 European countries. The network is disassortative with an assortativity coefficient of 0.1, with average degree $\langle k \rangle = 2.889$ and clustering coefficient of $C=0.127$. We model the system as a random resistor network carrying distributed flow and employ a capacity-limited overload model \cite{Motter_2002} to simulate cascading failures in the network (see Methods).
Note that despite the simplicity of this fundamental model for conserved flows (i.e. Kirchhoff's and Ohm's law in the resistor network), the underlying system of equations have {\em identical structure} to those of the DC power-flow approximation (the current and voltage corresponding to the power and phase, respectively) \cite{Gonzalez_2016,Zussman_2016,Korkali_arXiv}.

We find that the load is positively correlated with the degree, while the degree and load distributions span a relatively narrow range (see Suppl. Info. S2.), yet a significant variance of loads can be observed even for small degree values. This characteristic suggests that the load bearing responsibility of a node cannot be assessed exclusively from its degree. Using the spatial information of the nodes and edges, we plot the link length distribution (Suppl. Info. S2), and find that the majority of links span short distances with very few long range links constructed as part of the overall power grid designs. For the remainder of this article we use the terms removal of a node and attack on a node interchangeably to denote a node whose failure is used to initiate the cascade.

\subsection*{Random Capacity Allocation from Heterogeneous Distribution}

\begin{figure}[h!]
   \includegraphics[width=\textwidth]{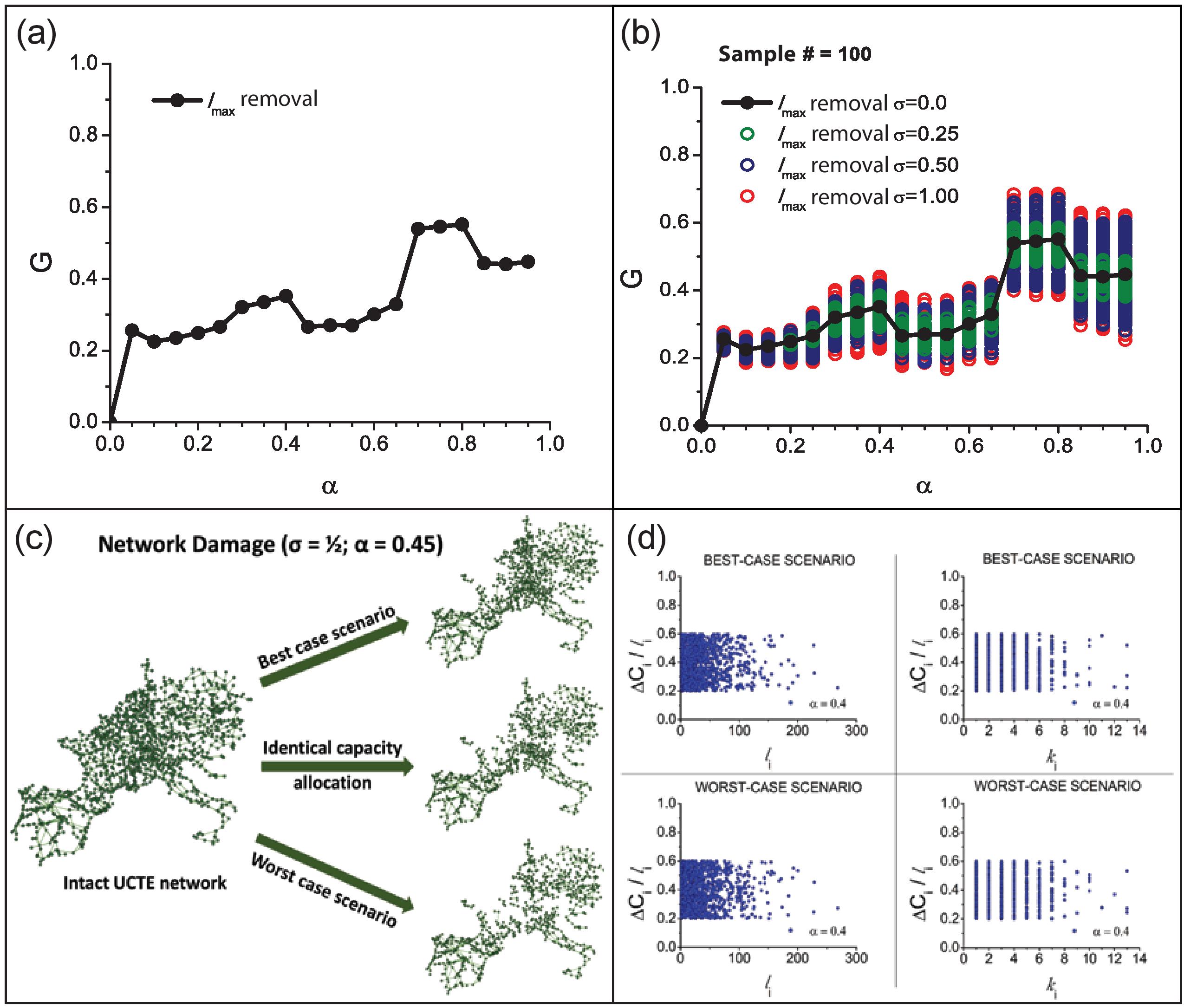}%
  \caption{Cascades on the UCTE network triggered by the removal of a single (highest load) node. (a) The cascade size as a function of tolerance parameter for (a) uniform capacity allocation; (b) stochastic capacity allocation of varying $\sigma$ values: $\sigma=0.25$ (green), $\sigma=0.50$ (blue), and $\sigma=1.00$ (red), with 100 simulations for each case. For comparison, the black connected data points show the performance of the uniform capacity allocation presented in (a). (c) Visualization of the damage caused in the UCTE power grid by the removal of the highest load node using the identical (uniform) capacity allocation vs. the best-case and worst-case scenario of the stochastic capacity allocations. (d) Correlation analysis of load/node degree vs. excess capacity for the best-case and worst-case scenario stochastic capacity allocations.}
  \label{fig-stochastic}
\end{figure}

We start studying the behavior of cascading failures in the UCTE network by assigning to each node a uniform capacity proportional to its load $C_i=(1+\alpha)l_{i}^{0}$, where $C_{i}$ is the total capacity of node $i$, $\alpha$ is a tolerance parameter \cite{Motter_2002}, and $l_{i}^{0}$ is the initial state load of node $i$, when the network is intact. Futher we will refer to $\Delta C_i =\alpha l_{i}^{0}$ as the excess capacity of node $i$. We study the cascading failure induced by the removal of a single (highest load) node from the network, and investigate the severity of the resulting damage, by calculating $G$, the size of the surviving giant component of the system.
In Fig.~\ref{fig-stochastic}(a) we present the size of the surviving giant component with respect to the tolerance parameter $\alpha$, and show that increasing $\alpha$ results in non-monotonic behavior of cascading failures. Higher capacity allocation (for example, an increase of $\alpha=0.4$ to $\alpha=0.45$) does not always reduce the severity of cascades, on the contrary, it may induce larger damage in the system \cite{Asztalos_2014}.

In real-world scenarios, allocating additional protection against failures in interconnected systems comes at a cost, and assigning higher capacity to each node in order to mitigate cascading failures in the power grid can be a costly procedure. Therefore, we introduce a different approach for allocating the resources (capacities) in the power grid by using a stochastic random capacity allocation method, while preserving the same total cost in the system as for the previous uniform capacity allocation case. The {\em mean} excess capacities assigned to the node $i$ are proportional to its normal equilibrium operational load $\overline{\Delta C_{i}}$$=$$\alpha {l_{i}^{o}}$. The particular values of the excess capacities, however, are drawn from a uniform distribution $\Delta C_{i}  \in  \alpha l_{i}^{0} [1 - \sigma, 1 + \sigma]$, with width $\sigma$ relative to the mean, such that $0 \leq \sigma \leq 1$, and subject to a fixed overall cost, $\sum_{i}\Delta C_{i}$$=$$\alpha \sum_i l^o_i$. In Fig.~\ref{fig-stochastic}(b) we show that our fixed-cost random node-capacity assignment results in a significant and emergent broadening of the distribution of the size of surviving giant component $G$, by varying $\sigma$, the width of the search space. Hence, we can stochastically search for optimal excess capacity allocations resulting in maximum resilience (largest $G$). We also find that by choosing higher $\sigma$ values, we can further improve the protection against damage with no additional cost (albeit with a saturating effect). For baseline comparison we also plot the initial case when identical relative excess capacity is assigned to each node (black filled symbols). The results clearly demonstrate that the fixed-cost stochastic distribution of resources (capacities) allows for identifying particular realizations which provide superior protection against cascading failures in the power grid.

In Fig.~\ref{fig-stochastic}(c) we illustrate for comparison the damage caused by the removal of the highest load node in case of uniform capacity allocation, the best-case scenario (the highest protection obtained from stochastic capacity allocation) and worst-case scenario (the lowest protection obtained from stochastic capacity allocation) when the tolerance is $\alpha=0.45$ and the width of the stochastic search space is $\sigma=1/2$. Aiming to better understand the properties of the capacity allocations resulting in the highest protection vs. the lowest protection for a given $\alpha$ tolerance parameter, we study in Fig.~\ref{fig-stochastic}(d) the correlation of the excess capacity with node degree and initial state load, but find little insight (based on traditional single-node structural or load-based characteristics) as to why a certain stochastic allocation of resources performs better than the others with respect to these node properties (see Suppl. Info. S3).

\subsection*{Sensitivity to Target Selection in Spatially-Localized Multi-Node Attacks}
Previously, we triggered cascades by removing a single (highest load) node. In order to capture more realistic scenarios, we consider a geometrically concentrated region containing several nodes, and analyze how the selection of multiple spatially-clustered initiators influences the size of the induced damage. Specifically, we investigate the sensitivity of the cascades to {\em small variations in the target set}. We arbitrarily choose the center of our attacks to be in South Western Europe. A circular region of radius $1r$, $2r$, and $3r$ contains 9, 12, and 20 nodes, respectively, where $r \approx 60 km$.
We consider various attack realizations (i.e., selections of target set) in the $2r$ and $3r$ regions while keeping the total number of attacked nodes {\em fixed} to be 9.

\begin{figure}[h!]
  \centering \includegraphics[width=0.88\textwidth]{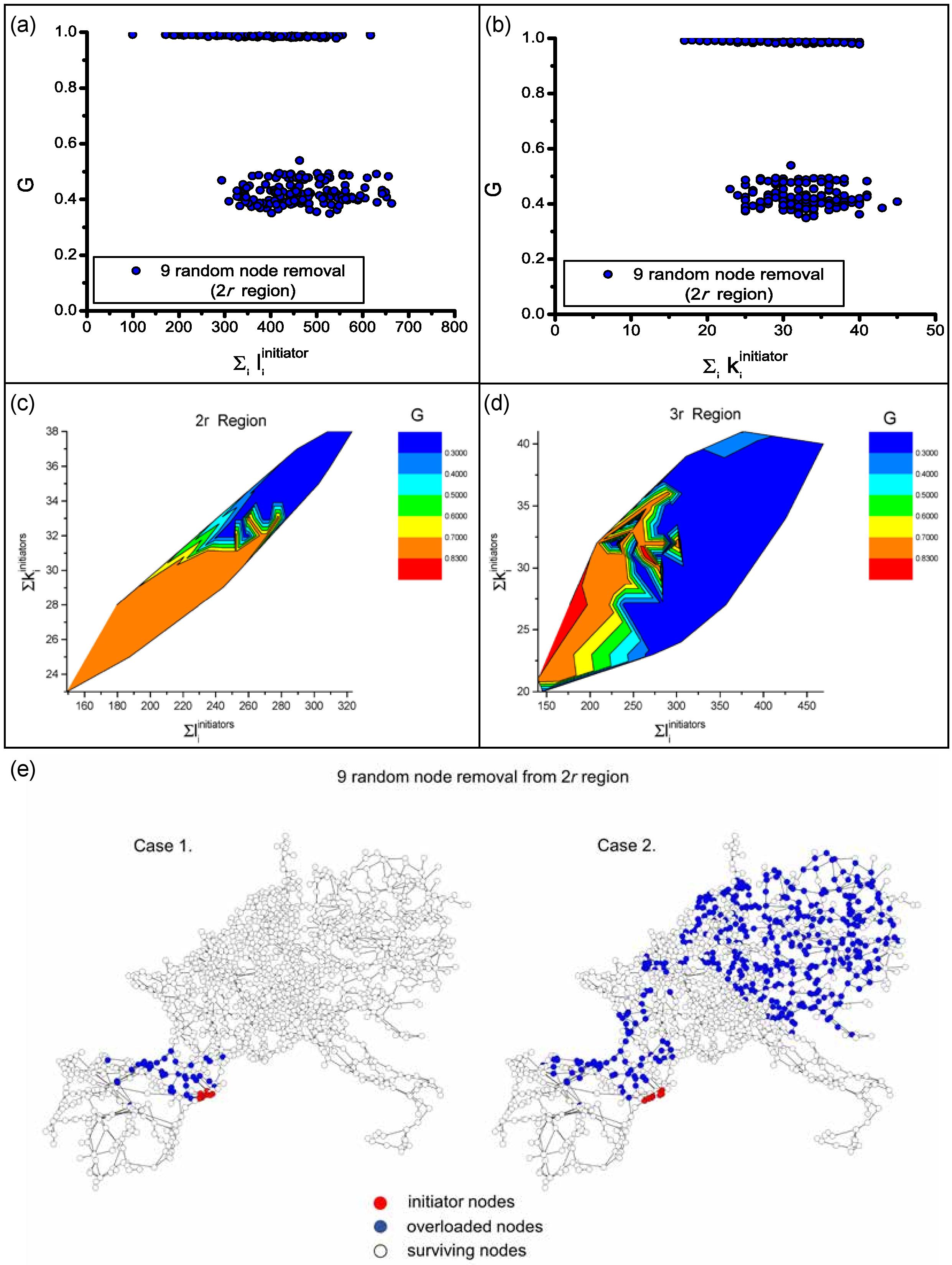}%
  \caption{Node sensitivity in the UCTE network. (a) $G$ vs. $\sum_{i} \ell_{i}^{initiator}$. (b) $G$ vs. $\sum_{i} k_{i}^{initiator}$. (c) Contour plot of 2$r$ region. (d) Contour plot of 3$r$ region. (e) Visualization of variability for small changes in the target set for spatially-clustered attacks on the UCTE network. Case 1 and Case 2 visualize the avalanche of failed nodes (blue) triggered by the removal of two distinct sets of 9 spatially localized nodes (red) from the same 2$r$ region, with $\alpha=0.1$}
  \label{fig-sensitivity}
\end{figure}

We have seen in the single node removal case a lack of correlation between the damage caused by a node, its degree, and its initial state load, respectively. Next, we analyze whether we can observe a correlation among these parameters in case of cascading failures triggered by multi-node attacks. In Fig.~\ref{fig-sensitivity}(a) we plot the size of the surviving giant component as a function of the sum of the initial state loads $\ell$ of the 9 initiator nodes removed from the 2$r$ region. Once again, we find that there is little or no correlation between traditional node characteristics (degree, load) and inflicted damage. Intuitively, one might expect to see a positive correlation between the load and damage, indicating that the failure of nodes with higher load would cause more severe damage. However, our results show that nodes with smaller load can cause severe damages, and nodes of similar load can induce minimal or no damage at all. Similarly, in Fig.~\ref{fig-sensitivity}(b) we see no correlation between the damage induced by the nodes and the sum of their degrees $k$. Despite the lack of correlations between these quantities, it is interesting to observe a clear {\em bimodal} behavior in Fig.~\ref{fig-sensitivity}(a),(b), with data points condensed around two regions: either causing severe damage of $G \in [0.3;0.6]$ or inducing no cascading failure. Similar bimodal behavior has been reported for cascading failures in the US Western Interconnect (USWI) power grid \cite{Buldyrev_2016}.

Expanding our analysis of multi-node failures to larger attack regions  (2$r$ and 3$r$),  we randomly select $9$ nodes to be the initiators for one simulation, and perform $100$ realizations for each region, and record the total degree and load of the initiators. In Fig.~\ref{fig-sensitivity}(c),(d) we demonstrate how the cascading process varies among different node degrees and loads. The two-level distribution of $G$ implies that for the safety of the entire system, it is better to lower the degree and load in a potentially targeted region. In the smaller $2r$ region, the low damage and high damage separates clearly and seems to be of the same size. It is easier to mitigate the damage of cascades in a smaller region if the targeted region has a smaller total degree and load. However, the area of high damage is much larger than that of small damage in the case of $3r$, meaning most realizations suffer high damage. The border of these two areas is not smooth, indicating that it is more complicated to reduce the damage than in the previous case. The results suggest that a larger region has a more complex topological structure, and that more variations of the initiators also increases the complexity of the resulting cascades.

For illustration, the nine nodes attacked and the resulting cascades for two different cases are shown in Fig.~\ref{fig-sensitivity}(e)). The visualizations clearly imply that there is a {\em very large variability} in cascade size for small changes in a spatially-confined target set, depending on which 9 nodes (indicated by red) are selected out of 12 within the $2r$ region. For Case 1, the resulting cascade (indicated by blue) is relatively closely confined to the attack region, while for Case 2, the cascade impacts a major portion of the network, including regions far away from the attack center.

The progression of the cascade initiated by the removal of different node sets chosen randomly from a spatially-concentrated region, as well as the progression of the cascade initiated by the removal of different node sets chosen randomly when enlarging the spatially-concentrated region (increasing radius $r$) can be viewed in the movies supplied as supplementary material (Movie S3-S11).

\subsection*{\textit{N}-Stable and \textit{(N-1)}-Stable Capacity Allocation for the UCTE network}

Next, we propose to study the UCTE network under two distinct system architectures based on power grid engineering methodology, namely the \textit{N}-stable and \textit{(N-1)}-stable configurations. The \textit{N}-stable UCTE system represents the power grid network in its intact form with capacity allocation $C_i=(1+\alpha)l_{i}^{0}$ \cite{Motter_2002}. Similarly, the \textit{(N-1)}-stable system is designed such that it is stable against any single node failure, when \textit{(N-1)} nodes are left in the network \cite{Bernstein_2011,Soltan_2014}. In order to attain this, we assign each node a capacity $C_{i}^{(N-1)}$, representing the minimum load capacity of node $i$ which guarantees that no subsequent overload failure occurs following any single node failure in the system. Based on this we express the capacity of node $i$ as $C_{i} = (1+\epsilon) C_{i}^{(N-1)}$, where $\epsilon$ is the relative tolerance. This method is used as an industry standard, and engineers commonly assign the relative tolerance values in the range of $\epsilon \in [0.1; 0.2]$. In our work we will further study the \textit{(N-1)}-stable using $\epsilon=0.10$. In order to effectively compare the efficacy of the two network construction methods, we calculate the $\alpha$ value of the \textit{N}-stable scenario that provides the same cost of protection as the \textit{(N-1)}-stable scenario with $\epsilon=0.10$. We find that $\alpha=0.5927$ is the value for which the two systems use the same amount of resources, thus in the following analyses we will use  $\alpha=0.5927$ tolerance parameter for the \textit{N}-stable case, and  $\epsilon=0.10$ relative tolerance for the \textit{(N-1)}-stable case, unless otherwise stated.

\begin{figure}[h!]
   \includegraphics[width=\textwidth]{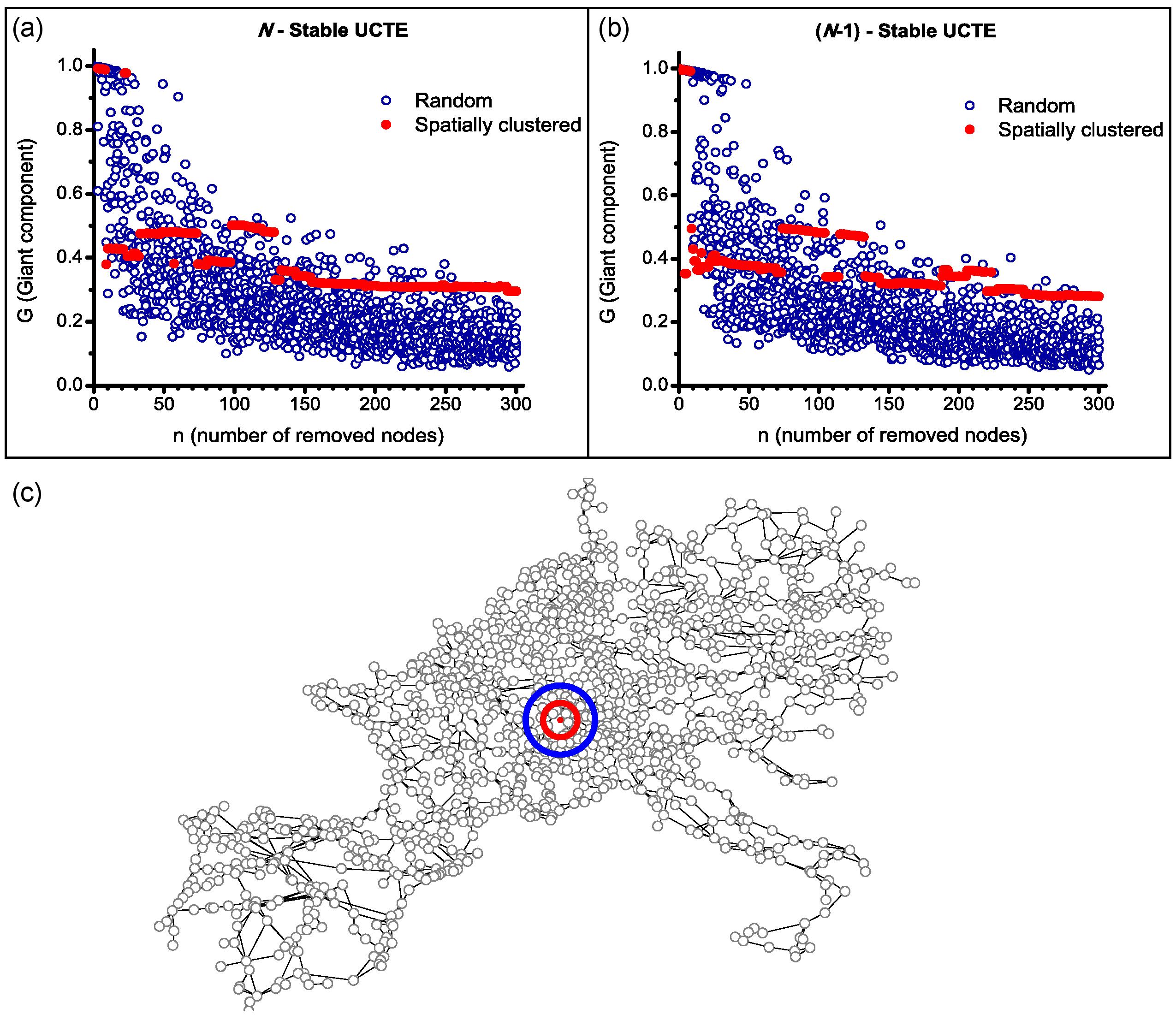}%
  \caption{Spatially-concentrated and distributed (random) attacks in \textit{N}-stable and (\textit{N}-1)-stable UCTE networks. For each $n$ there are 5 realizations of random attacks. The tolerance parameter of the \textit{N}-stable system is $\alpha = 0.5927$, and for the \textit{(N-1)}-stable $\epsilon = 0.1$. (c) Visualization of $1r$ (red circle) and $2r$ (blue circle) regions used for spatially localized attacks in the UCTE network, where the distance is calculated from the geographic center of the network (red dot).}
  \label{fig-comp}
\end{figure}

Of great interest when studying attacks on networks is the role that spatial correlations, as well as the initial capacities on the nodes or edges of the intact network play in determining the severity of a cascade. Fig.~\ref{fig-comp}(a),(b) show the damage delivered by attacks on \textit{N}-stable and \textit{(N-1)}-stable networks, respectively. To initiate the cascade, nodes were either randomly chosen (with 5 realizations for each value of $n$), or removed in succession of distance from the center of the network (spatially correlated). The horizontal and vertical axes represent the number of removed nodes ($n$) triggering the cascade and the size of the surviving giant component ($G$) after the cascade, respectively. In both figures, the sum of the allocated capacities are equivalent, giving a fixed-cost comparison of the two system designs. It is immediately apparent that for small $n$ ($n<100$) the \textit{N}-stable and \textit{(N-1)}-stable configurations are more stable against random failures, but are more vulnerable to spatially-localized attacks. As the number of targeted nodes becomes larger ($n\geq100$), both systems become more vulnerable against random failures and more robust against localized attacks. What may be the most striking are the discontinuities in $G$ for the spatially-correlated scenario. As the radius of the attack increases, the severity of the damage slowly changes until certain critical radii are reached. There we witness abrupt changes in $G$, making the network more, or less robust depending on the location. This suggests that the severity of an attack on a network varies {\em non-monotonically} with the size of the attack. Smaller attacks may yield more damage and vice versa, similar to our previous findings. Also of importance is the fact that for smaller attacks, the spatially correlated attack is much more effective than the random attack. Possible reasoning for this observation is that a small random attack is unlikely to strike pivotal nodes, while a spatially correlated attack has a much larger impact on one local region, disrupting the network more severely. The opposite is observed when the attack is large, and along similar lines of thought, it is unlikely that pivotal nodes in the network are clustered in one region. By dispersing the attack, the probability of hitting a higher number of such nodes increases.

Fig.~\ref{fig-comp}(c) highlights the 1$r$ (red circle) and 2$r$ (blue circle) regions calculated from the geographic center of the UCTE network. In further spatially-localized attacks we will use these regions to analyze cascades induced by clustered node failures.

\begin{figure}[h!]
  \includegraphics[width=\textwidth]{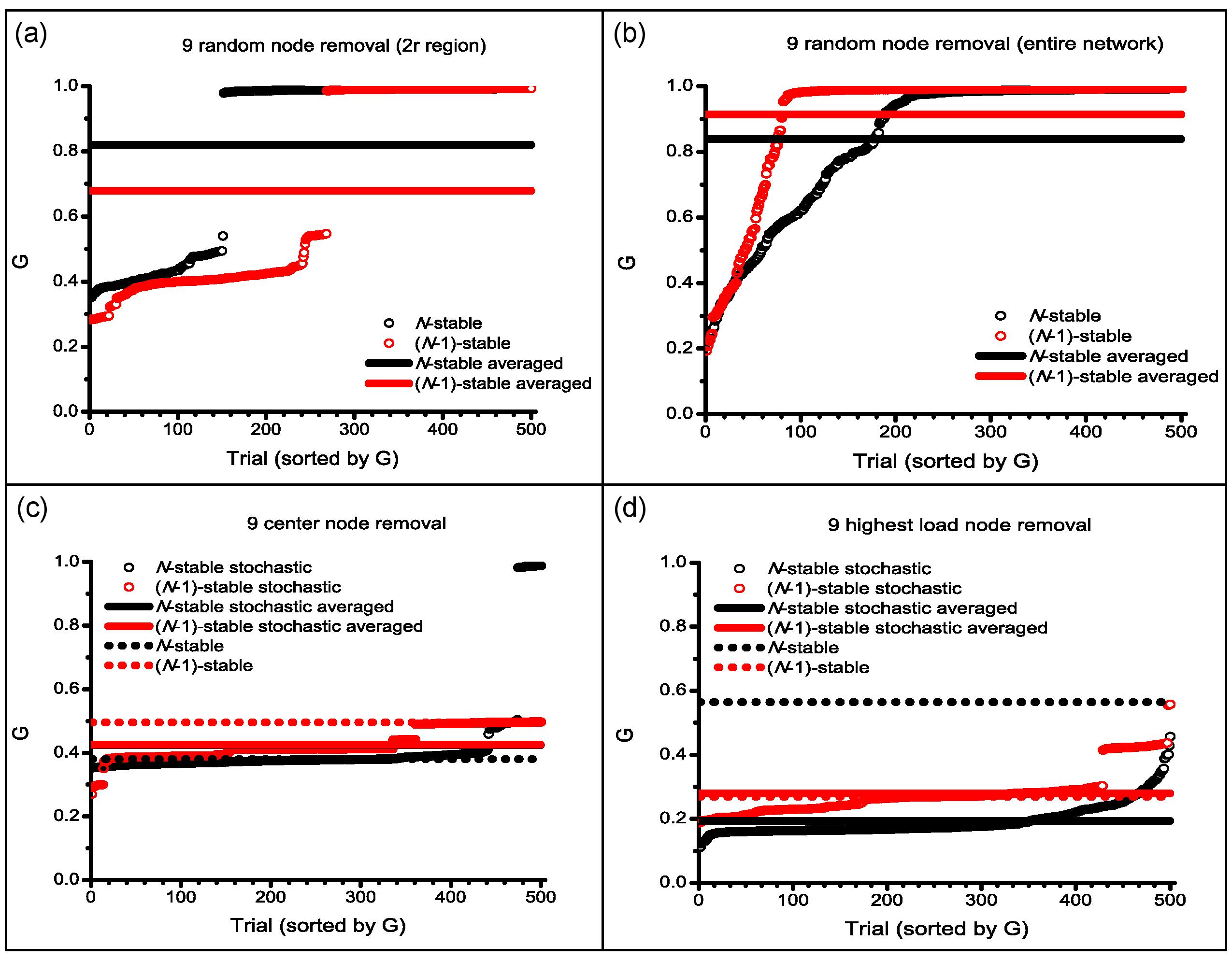}%
 \caption{Multi-node attacks on \textit{N}-stable and \textit{(N-1)}-stable UCTE networks. (a) 9 random node removal from the 2$r$ region; (b) 9 random node removal from the entire network; (c) 9 center node removal with stochastic capacity allocation;  (d) 9 highest load node removal with stochastic capacity allocation. Each removal scenario contains 500 repetitions on \textit{N}-stable ($\alpha=0.5927$) and on \textit{(N-1)}-stable ($\epsilon=0.1$) UCTE networks. For clarity purposes the sizes of the surviving giant component recorded at each trial are sorted in non-decreasing order.}
  \label{fig-9rem}
\end{figure}

Next, we study the performance of the \textit{N}-stable and \textit{(N-1)}-stable UCTE network against various multi-node attacks. We remove 9 nodes from a 2$r$ region, from the entire network, from the center, and finally, the 9 highest load nodes, and record the size of the surviving giant component $G$ for each scenario. For clarity, we sort the simulations in non-decreasing order of damage $G$ and present our results in Fig.~\ref{fig-9rem}. In Fig.~\ref{fig-9rem}(a), we again see the bimodal nature of spatially-localized attacks, either causing severe damage or not triggering cascading failures. Using the averaged values of the 500 simulations (solid lines), we can compare the performance of the two UCTE system designs, to conclude that for spatially-localized attacks in a 2$r$ region the \textit{N}-stable configuration outperforms the \textit{(N-1)}-stable network. For the removal of 9 spatially-localized nodes we see that the \textit{N}-stable scenario triggers cascading failures in less than 25\% of simulations, whereas in the \textit{(N-1)}-stable network, cascading failures are induced in approximately 50\% of cases. Similarly, in the case of removing the 9 highest load nodes, we find that the \textit{N}-stable configuration is almost 30\% more robust than the \textit{(N-1)}-stable network (see Fig.~\ref{fig-9rem}(d), dotted lines).
In Fig.~\ref{fig-9rem}(b) we can see that for 9 random node removals (from the entire network) the robustness of both systems is improved, however in this case the \textit{(N-1)}-stable configuration performs better, by inducing cascading failures in less than 20\% of the simulations. Likewise, in the case of removing 9 center nodes from the network, the \textit{(N-1)}-stable configuration is more robust than its \textit{N}-stable counterpart (see Fig.~\ref{fig-9rem}(c), dotted lines).

Using the stochastic capacity allocation method employed in the case of single node removal, we study the damage caused by multi-node attacks when the \textit{N}-stable and \textit{(N-1)}-stable networks are assigned heterogeneous excess capacities. In the case of the \textit{(N-1)}-stable configuration, the excess capacities are assigned from
$\Delta C_{i} \in \epsilon C_{i}^{(N-1)} [1 - \sigma, 1 + \sigma]$,
mean relative tolerance $\epsilon=0.10$, search width $\sigma=1$. We find that in the case of 9 center node removal (Fig.~\ref{fig-9rem}(c)), using stochastic capacity allocations, we can improve the performance of the \textit{N}-stable configuration, while maintaining the same fixed-cost. It is important to notice that we can find optimal stochastic capacity allocations such that the 9 center node removal does not induce cascading failures, maintaining the \textit{N}-stable network intact. Also, as we have seen in the case of single node failure, with stochastic capacity allocation we can also obtain allocation scenarios that offer higher protection than the uniform capacity allocation case. For the \textit{(N-1)}-stable network, however, we cannot find stochastic allocation of resources that can outperform the uniform one. Moreover, we can see that each stochastic capacity allocation sample makes the system more vulnerable to failures. Since the \textit{(N-1)}-stable configuration is already optimized for any single node failure, the stochastic capacity allocation perturbes this optimization, thus resulting in a more vulnerable system. In contrast, for the case of 9 highest load node removal, Fig.~\ref{fig-9rem}(d) shows that we can find random capacity allocations that improve the robustness of the \textit{(N-1)}-stable system by almost 30\%. In addition, we can see that, on average, the \textit{(N-1)}-stable configuration with random capacity assignment only slightly improves robustness of the \textit{(N-1)}-stable system with uniform capacity allocation, however, it outperforms the \textit{N}-stable network with random capacity assignment. Finally, overall, we can see that the \textit{N}-stable network with uniform capacity allocation still remains significantly more robust than stochastic capacity allocations on both network constructs.

\subsection*{Multi-node Attack Strategies}

\begin{figure}[h!]
   \includegraphics[width=\textwidth]{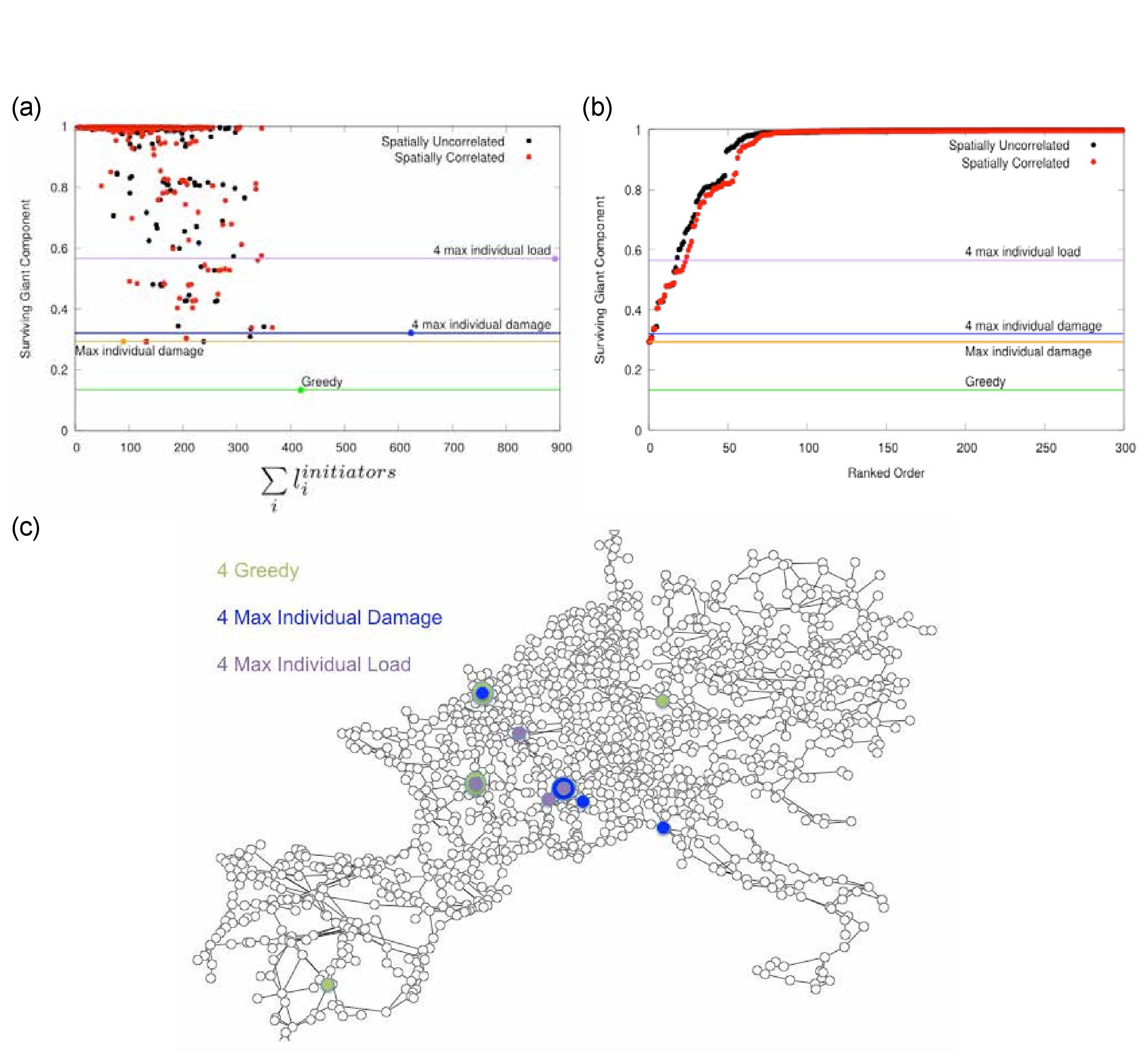}%
  \caption{Severe attack patterns in the \textit{N}-stable UCTE network. (a) Size of the surviving giant component $G$ as a function of the total load of targeted nodes. The data points on the continuous lines represent the sizes of the total load of the targeted nodes; the lines show the sizes of $G$ when targeting those nodes. Their color codes identify the used attack pattern: the 4 highest load node removal (purple), the 4 nodes that removed together cause the highest damage, nodes selected by our greedy algorithm (green), the 4 nodes causing the highest damage when individually removed (blue). (b) Comparison of the size of damage caused by various attack patterns by ranking $G$ in nondecreasing order. The figures contain 300 simulations for each attack scenario on the \textit{N}-stable UCTE network with $\alpha$ = 0.5927. (c) Visualization on the UCTE network of the nodes used in the different attack patterns for triggering the cascading failures.}
  \label{fig-attack}
\end{figure}

Aiming to capture the consequences of various targeted attack strategies or random failures in power grids, we further expand our analysis, and study five different damage scenarios in the UCTE network that can truthfully reproduce possible attack schemes or breakdowns encountered in real-world situations. In the first method (random) we randomly select 4 nodes from the network and remove them simultaneously, thus depicting the outcomes of multi-node random failures or random attacks on the network. In the second method (clustered) we remove 4 spatially concentrated nodes, simulating cascading failure in case of an attack on a single large region. In the third method (4 max. individual damage) we select four nodes from the network that individually generate the most severe cascades, and remove them simultaneously. Next, the 4 max. individual load method reflects the scenario of simultaneously attacking the highest load nodes in the network. We can observe a fundamental non-monotonic feature of the cascading overload failure process: attacking only the (single) largest-damage inducing node in the network results in a smaller surviving giant component $G$ (and a larger overall damage) than the {\em simultaneous} attack of the 4 max. individual load nodes [Fig.~\ref{fig-attack}(a)].

Lastly, in the fifth method (greedy) we select the 4 nodes that trigger the largest cascading failures when removed simultaneously. In order to find these nodes, first we identify the single most damaging node by scanning all $N$ single node failures. Then we determine the second node that causes the highest damage by scanning all remaining \textit{N}-1 nodes on top of the failure caused by the first identified node, and continue in this manner until we have identified a set of the 4 most damaging nodes. The comparison of the effects of these five cascade triggering methods is captured in Fig. \ref{fig-attack}(a). The data points comprise 300 runs for each scenario on the \textit{N}-stable UCTE network, with tolerance parameter $\alpha = 0.5927$. We find that the random and clustered attacks cause similar damage in the network, and in most cases the system remains intact, without triggering a cascading failure. In Fig. \ref{fig-attack}(b) we rank the size of the damage ($G$) in non-decreasing order to create a more visually compelling comparison of the impact of the damages induced by the different attack scenarios. This figure shows that among 300 simulated random and spatially clustered attacks, only about 50 (approx. 15 \%) cases induce damage in the system. However, attacks by the three other selection strategies trigger more severe cascading failures in the power grid. Moreover, the greedy method, where we choose the 4 most damaging nodes, triggers the strongest cascading failure, under which the surviving giant component is nearly 10\% of the initial network. This means that removing only 4 nodes can induce an almost complete system failure, a magnitude of damage that no other method can achieve. On the other hand, the random node failures or spatially-clustered attacks have minor effects on the network, causing the smallest damage in the system compared to the other attack strategies.

\subsection*{Phase Transition in Cascading Failures}

\begin{figure}[h!]
  \includegraphics[width=\textwidth]{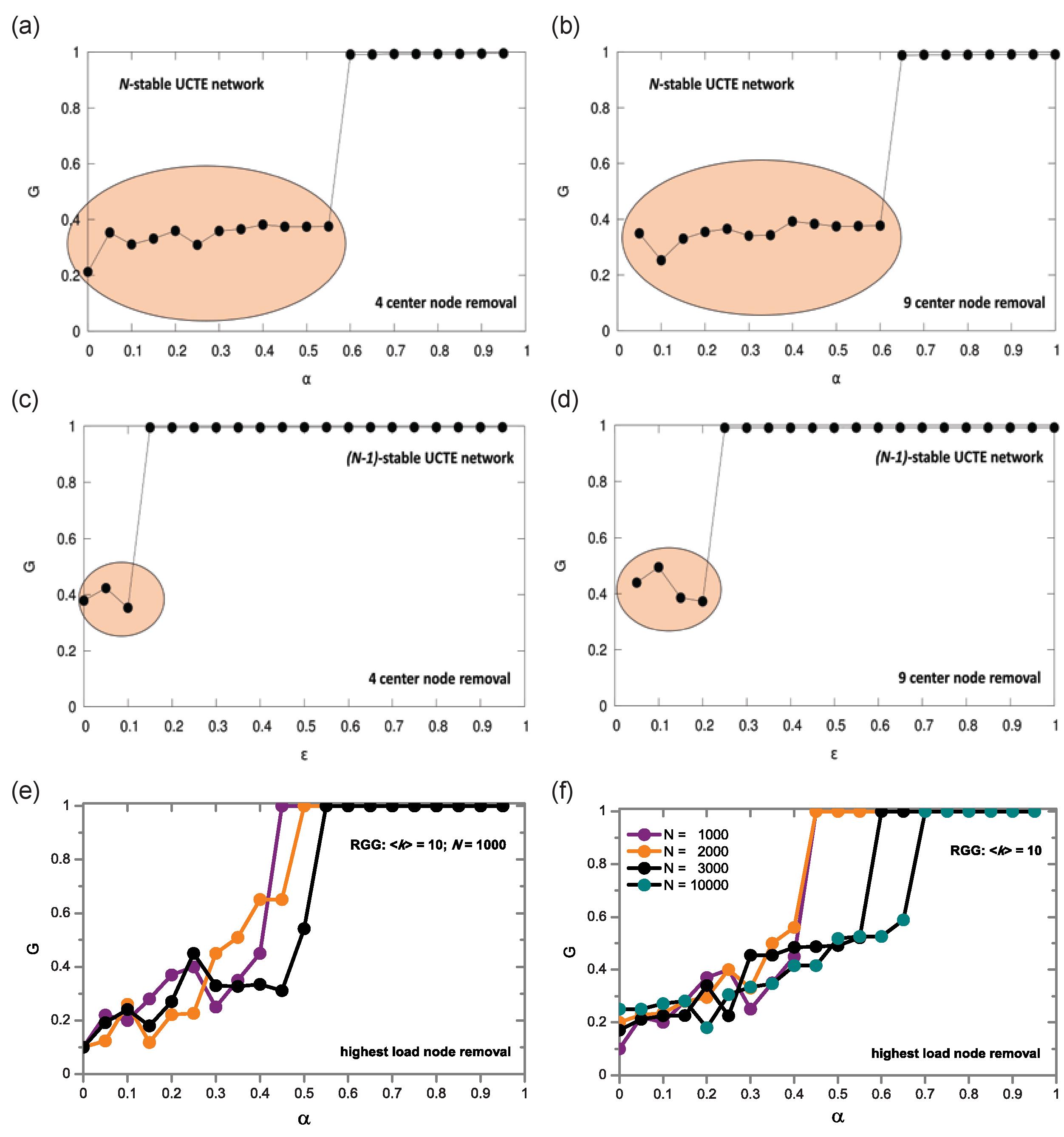}%
  \caption{Phase transitions with increasing protection in UCTE and RGG networks. (a) 4 center node removal in \textit{N}-stable UCTE network. (b) 9 center node removal in \textit{(N-1)}-stable UCTE network. (c) 4 center node removal in \textit{N}-stable UCTE network. (d) 9 center node removal in \textit{(N-1)}-stable UCTE network. The pink highlighted areas depict the region before the transition, after which the network becomes fully protected against failures. Phase transition in RGGs when cascades are triggered by the removal of the highest load node (e) with 3 different RGG network samples of identical size $N=1000$ and average degree $\langle k \rangle=10$; (f) with 4 realizations of RGG of same average degree $\langle k \rangle=10$, but different system sizes.}
  \label{fig-phase}
\end{figure}

Asztalos et al. \cite{Asztalos_2014} have reported that by increasing the tolerance parameter in the system, one can observe a non-monotonic behavior in the size of the surviving giant component. Likewise, as seen in Fig.~\ref{fig-phase}(a)-(d), we show that in the \textit{N}-stable and \textit{(N-1)}-stable UCTE network configurations, by varying the value of $\alpha$ and $\epsilon$ tolerance, respectively, we can also notice a similar, non-monotonic trend for various spatially localized attacks. The encircled highlighted regions depict the tolerance values for which the system becomes vulnerable to large-scale cascading failures. By increasing the protection beyond these values, in both \textit{N}-stable and \textit{(N-1)}-stable cases, we can determine a certain tolerance value, where the system undergoes a phase transition, after which point the network almost completely survives the attack. The \textit{(N-1)}-stable network reaches full protection at a much smaller tolerance value, due to the fact that it is already designed to be protected against any single node failure, thus it is also more robust against multi-node failures.

In addition, aiming to provide a generalized behavior in spatial networks of the observed transition property, we study the phase transition of cascading failures in RGGs, and find that despite using a different removal strategy, single (highest load) node removal, we obtain similar bimodal behavior in the size of induced damages. However, as Fig.~\ref{fig-phase}(e) indicates, we cannot assess the tolerance value where the phase transition occurs, because different spatial network instances, despite having identical network properties (system size, average degree), present a very different non-monotonic increase with critical points at different tolerance values. Moreover, we show in Fig.~\ref{fig-phase}(f) that by increasing the system size, we can see the same non-monotonic behavior in the severity of cascades, with a phase transition occurring at different critical $\alpha$ values. Once again these results indicate the complex nature of spatial networks, and the difficulty in predicting the behavior of cascading failures in such systems.

\subsection*{Cascade-size Distribution Analysis}

\begin{figure}[h!]
  \includegraphics[width=\textwidth]{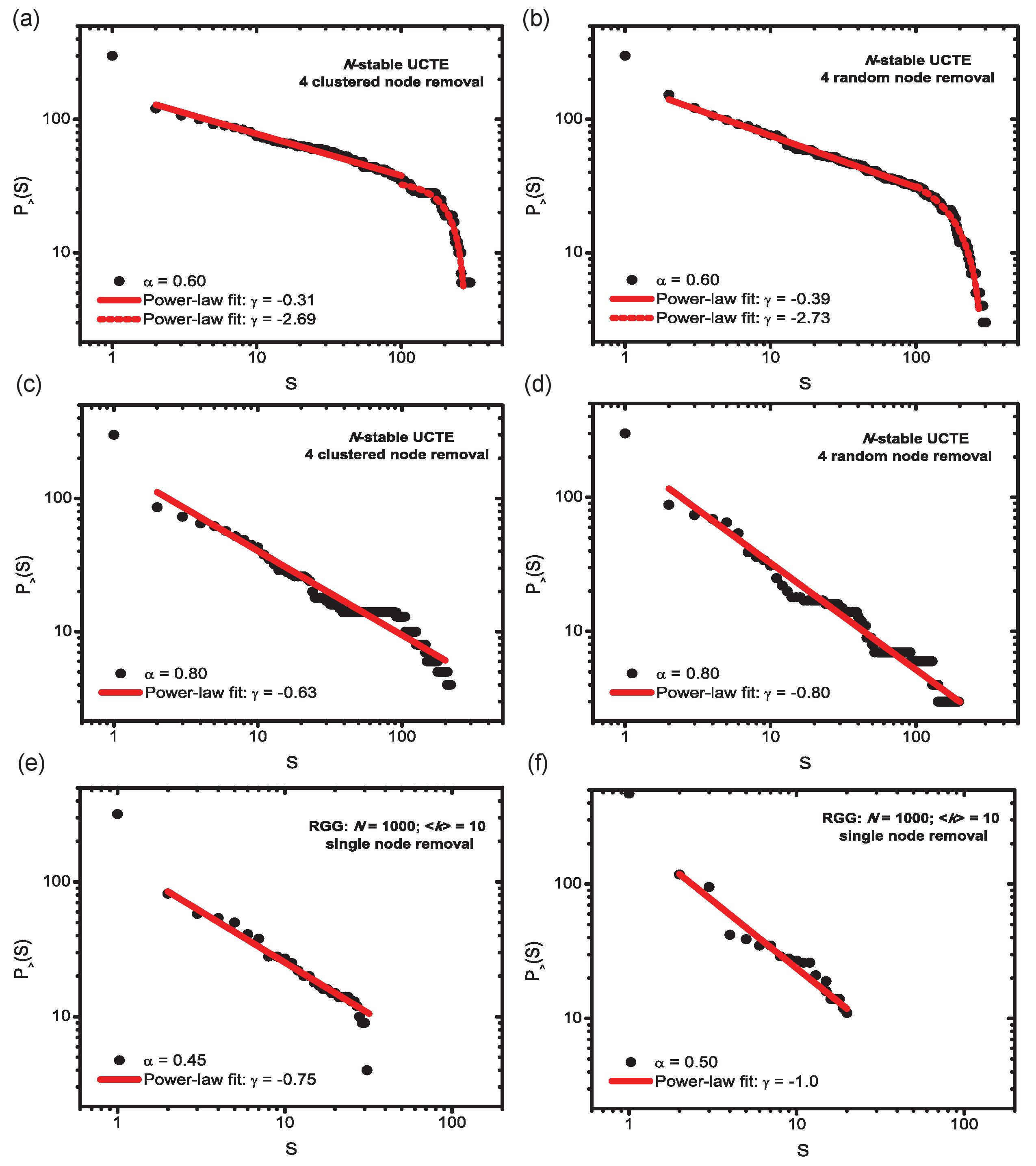}%
  \caption{Power-law tail of cascade size distributions in the phase transition regime. (a) Power-law fit (red line) of clustered 4 node removals (black dots) with $\alpha = 0.6$. (b) Power-law fit (red line) of random 4 node removals (black dots) with $\alpha = 0.6$. (c) Power-law fit of clustered 4 node removals (black dots) with $\alpha = 0.8$. (d) Power-law fit (red line) of random 4 node removals (black dots) with $\alpha = 0.8$. Small event regime $1 \leq S \leq 200$, large event regime: $201 \leq S \leq 300$.}
  \label{fig-pl}
\end{figure}

Next, we study the cascade size distributions in the phase transition regimes, where, according to prior studies, such distributions are characterized by power-law tails. We quantify the cascade size $S$ as the number of nodes that fail during a cascading process. For comparison, in Fig.~\ref{fig-pl} we plot the cascade distributions both as $G$ and $S$. Since $S$ is the number of failed nodes, and $G$ is the size of the surviving giant component, $S \neq N-G$. However, in Suppl. Info. S9 we demonstrate that there is a clear linear correlation between these two quantities, and plotting data as a function of $G$ or $S$ provides similar results. The cumulative cascade size distributions exhibit a power-law tail $P_{>}(S) \sim \frac{1}{S^{\gamma}}$ with exponent $\gamma$, and cascade size distributions can be found as $p(S) \sim \frac{1}{S^{\gamma+1}}$.  We focus on cascading failures on the \textit{N}-stable configuration triggered by spatially clustered sets of 4 nodes or random sets of 4 nodes distributed throughout the network for an $\alpha$ value fixed within the phase transition region, and record the distribution of cascade sizes as the fraction of blackouts of size $S$ or larger. In Fig.~\ref{fig-pl}(a) and Fig.~\ref{fig-pl}(b) we show that both triggering methods generate power-law tail distributions. We also find that increasing the $\alpha$ tolerance parameter from $\alpha=0.60$ to $\alpha=0.80$ the cascade size distribution becomes more abrupt, producing a higher power-law exponent. This observation can be explained intuitively; by increasing the tolerance parameter, we increase the protection in the system, and reduce the number of large cascades, thus increasing the number of small failures. As seen in these figures, for $\alpha=0.60$ both cascade size distributions triggered by clustered sets of 4 nodes and random sets of 4 nodes exhibit a dual plateau. Thus, in Fig.~\ref{fig-pl}(a),(b) we analyzed these as two separate event regimes: the small event regime, where the size of the cascading failures is small $(1 \leq S \leq 200)$, and the large event regime, where the resulted failures are large $(201 \leq S)$. We can see that both triggering methods produce similar cascade size distributions, with power-law exponents of $\gamma \approx 0.3$ in the small event regime, and $\gamma \approx 2.7$ in the large event regime. Our results are in agreement with previous work \cite{Kertesz_2006}, where the probability density of conductance changes also follow a power-law, with two different regimes, reported both for synthetic networks and the Norway power grid system. Moreover, the values of the reported power-law exponents in the two distinct regimes are close to the values observed in our work on the UCTE network. We employ statistical testing methods to assess the validity of power-law fitting, and describe the used methodology briefly in the Methods section and in detail in Suppl. Info. S9.
We also show in Suppl. Info. S9 that the power-law distribution fails to occur when the value of $\alpha$ and $\epsilon$ tolerance parameters are not within the phase transition, which is in agreement with the observation reported in \cite{Lee}. In addition, in Fig.~\ref{fig-pl}(e),(f), we test our finding on random geometric graphs (RGGs), and find a similar power-law trend in the cascade size distribution when we trigger the cascade by the removal of a single (highest load) node. Also in RGGs, as we move away from the phase transition region, the power-law characteristic vanishes (see Suppl. Info. S9).

\subsection*{Predicting the Severity of Cascading Failures}

\begin{figure}[h!]
  \centering\includegraphics[width=0.98\textwidth]{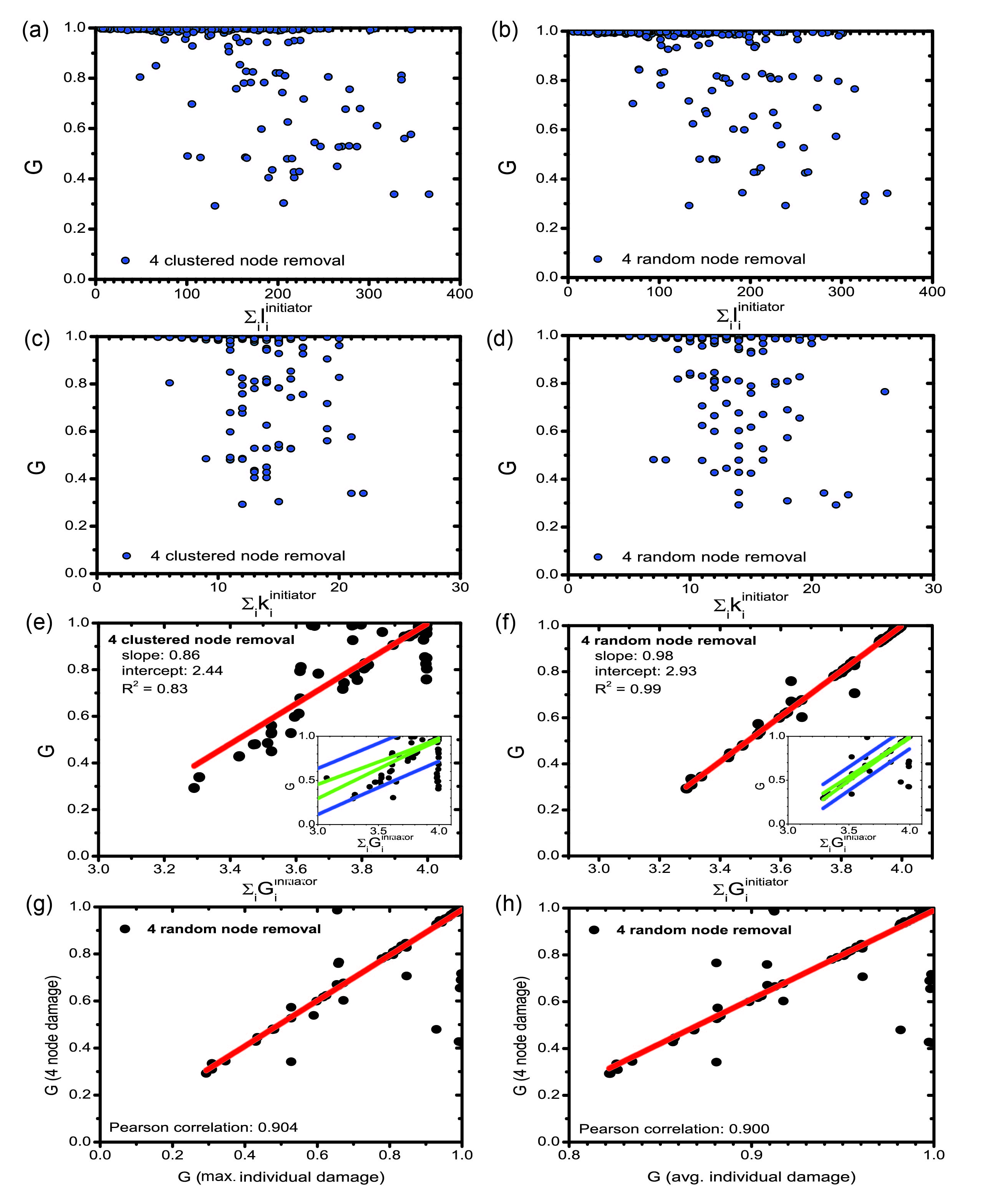}%
  \caption{Correlation analysis of failures triggered by the removal of 4 nodes. $\alpha$ = 0.5927 (relative tolerance); 300 repetitions of 4 random node removals from the entire network.}
 \label{fig-corr}
\end{figure}

So far we have analyzed mitigation strategies and the behavior of cascading failures triggered by various strategies of node removals. Here, we study the predictability of damage caused by the failure of a set of nodes $G^{All initiators}$, given the sum of damages caused by each individual node failure, $\sum G_{i}^{initiator}$ which are the sums of the surviving giant components (in Suppl. Info. S10 we also present this analysis for $S$ avalanche sizes). Using various numbers of initiator nodes and different removal strategies, we show in Fig.~\ref{fig-corr}(e),(f) that there is a strong linear correlation between the variables. This indicates that given a set of node failures, and knowing the damage induced solely by the failure of each individual node, we can effectively predict the damage induced in the event of a multi-node failure.
The insets show results for 300 samples, however, based on the confidence intervals of the data, we have removed a small fraction of outlier nodes. A detailed description of the strategy used to identify these outliers is presented in the Methods section and Suppl. Info. S10.
Additionally, in Suppl. Info. S10 we present these correlations for different number of node removals from 2 up to 9, and show that as we increase the number of removed nodes, the correlation significantly drops. This finding once again suggests that accurate estimation of damage is harder to attain as the complexity of the cascade (number of triggering nodes) increases. 

To understand the mechanism responsible for the correlation between single node and multiple node attacks, we investigate various characteristics of the cascading failures in question. We find that the sum of individual node failures $\sum G_{i}^{initiator}$ and the most damaged individual node $\min G_i^{initiator}$ (lower G represents higher damage) correlate equally with a multi-node failure $G^{All initiators}$ (Suppl. Info. S10). This suggests that for multi-node failure scenarios, the node which individually inflicts the highest level of damage on the network is responsible for the progression of the multi-node cascade, which implies the existence of dominating nodes of cascading failures. For most scenarios, the remaining nodes have little to no effect on the progression of the cascade. However, when multiple nodes, that individually inflict the highest damage, are triggered simultaneously; we find a poor correlation between iindividual and multi-node failures. 

Finally, in order to study whether there is a correlation between initial network characteristics and cascade severity, we run extensive numerical simulations of individual node removals for varying $\alpha$ values. The correlation over all attacks for various alpha values are shown in Suppl. Info. 10. We find that as the tolerance of the network is increased, the system remains unaffected by a higher percentage of the attacks. A correlation over various alpha values would indicate that the network topology is responsible for the cascade progression. We find a weak, but existent correlation suggesting that while topology is important, the cascade progression is also highly dependent on tolerance value. This finding is intuitive, since the network topology is continually disintegrated throughout the cascading failure process.

\section*{Discussion}
We have studied how cascading failures induced by load redistribution caused by single- and multi-node failures on spatial networks carrying distributed flow propagate through the network. We conducted our research on a real-world power transmission network and in spatially-constrained synthetic network model, RGGs. We found that one cannot assess the severity of a damage induced by single- or multi-node attacks based on the degree or initial state load of the initiator nodes.

Furthermore, we have demonstrated that small ($n<100$) spatially-localized attacks or failures cause a higher damage than spatially distributed, random failures. Our results indicate that intentional attacks on the network cause a more severe damage, however, providing protection to the system against particular attack scenarios results in a more vulnerable system against other unexpected failures. On the other hand, when the system experiences a massive node failure ($n\geq100$), the distributed attacks prove to be more damaging than the spatially-localized ones, suggesting that an extended random attack is more likely to find critical nodes that can induce severe damages.

We have proposed a fixed-cost stochastic capacity allocation strategy for mitigating single-node failures. In addition, we designed the \textit{(N-1)}-stable UCTE network that remains stable in the event of any single-node failure. We demonstrated that while these methods can efficiently mitigate against single-node damage, they offer little protection in case of cascading failures induced by multi-node attacks. Specifically, for typical values of node-level tolerances $(\epsilon \approx 10-20\%)$, \textit{(N-1)}-stable networks can still develop large cascades for both spatially-concentrated and distributed multi-node attacks. In addition, we have shown that single node characteristics (load, degree) offer little insight on the severity of induced cascades, and it is not apparent on which nodes one should increase protection to successfully mitigate cascading failures.

In Fig.~\ref{fig-sensitivity}(e) and Movie S3-S11 we have exposed the non-local nature of the spread of cascades. We have seen that triggering a failure in a localized region does not necessarily induce failures in the surrounding area. In contrast, cascading failures can propagate throughout the entire network and cause failures in regions found at large spatial distances from the targeted area. Moreover, we have shown that any small change in the target set can cause a large variability in the size of induced damage.

Analyzing the survivability of the system by increasing the excess capacity, we have revealed the non-monotonic behavior in both \textit{N}- and \textit{(N-1)}-stable UCTE networks, and also in RGGs. 

Additionally, we have shown that even for large system size RGG networks, non-self averaging prevails; for large ($10^4$) network size the system still exhibits a non-monotonic fluctuating behavior in $G$ vs $\alpha$. We have also found non-monotonic behavior when analyzing the size of the surviving giant component by increasing the number of targeted nodes. We have seen that cascade sizes exhibit bimodal nature characterized by a first-order phase transition; in the critical regime, cascade-size distributions follow power-law trends, behavior that vanishes as we move away from the critical regions. Lastly, we have demonstrated that one can predict with high accuracy the damage triggered by multi-node attacks, if the damage size that each node can trigger individually is known. However, we have also shown that as the complexity of the induced damage (number of failed nodes) increases, the predictability becomes ineffective. We have also shown that even if we maintain the same number of initiator nodes, by enlarging the geographic area on which the attack is triggered, we obtain high variation in the size of induced damage. This indicates that the size of the region in which the failure can occur also adds to the complexity of the cascading failures, and hardens mitigation strategies in case of a fixed number of node attacks or random failures in expanded regions.

In summary, we have found that traditional single-node measures, frequently employed in network science (such as degree or load), are largely ineffective for predicting sensitivity and severity of cascades or for mitigating against them. The cascade behavior also exhibits fundamental {\em non-monotonic} features, very different \cite{Bernstein_2011,Asztalos_2014} from cascades in epidemic spreading or social contagion: ({\it i}) a larger number of nodes initially attacked may lead to smaller cascades; ({\it ii}) increasing the node (or edge-level) tolerance uniformly across the network may lead to larger cascades, i.e., indiscriminately investing resources in the protection of nodes or links can actually make the network more vulnerable against cascading failures (``paying more can result in less", in terms of robustness). Both of these issues are implicitly (but inherently) related to existence of ``fuse" lines and nodes in the network, whose identification is a computationally hard problem. We also observed that cascading failures are {\em non-self-averaging} in spatial graphs, hence predictability is poor and conventional mitigation strategies are ineffective. Further, in part related to the above, we have shown that cascade sizes exhibit {\em large variability for small changes in the target} set. It is also important to note that while the spatial propagation of the cascading failures resembles wave-like propagation features (overloads farther away from the original attack occur later) \cite{Asztalos_2014,Havlin_2016}, some {\em non-local} features are apparent of our cascade visualizations. The overload failure process is {\em not} a nearest-neighbor spreading process in the grid (by virtue of the current flow equations and the possible non-local redistribution of overloads); regions along the path of the cascade can be spared or "bypassed", while possibly giving rise to large-area outages far away from the location of the original attack.

In conclusion, we have used extensive real-world targeted node attack and random node failure scenarios to induce cascading failures on spatial networks, and analyzed their characteristics and propagation. We have introduced effective mitigation strategies to reduce the severity of damages, and presented the limits of predictability of assessing the severity of disruptions in the system, due to the complex nature of cascading failures occurring in spatially-embedded networks.

\section*{Methods}
Below we describe the distributed flow model and cascade model that we use in our study.

\subsection*{Distributed flow}
We assume that the flow is distributed, directed and of unit size, associated with a source and sink, and flow through all possible paths between source and sink. We model the network as a simple random resistor network with unit conductances along the edges \cite{Asztalos_2012,Korniss}. In this model each node and edge is involved in transporting current from source to sink, therefore each link experiences a load which is the current along that edge. For a link connecting nodes $i$ and $j$ the load is calculated as $\ell_{ij}=I_{ij}^{st}$, and the load on an arbitrary node $i$ is the net current flowing through that node $\ell_{i}=I_{i}^{st}$. The two loads can be expressed as

\begin{equation}
I_{i}^{(st)} = \frac{1}{2} \sum_{j} |I_{ij}^{(st)}|
\end{equation}

Next, we assume that all nodes are simultaneously sources and for each source we randomly choose a target from the remaning $N-1$ nodes. Thus, we assume that unit current flows simultaneously between $N$ source/target pairs, and the load is defined as the superposition of all currents flowing through an arbitrary node. This is identical to the node current-flow betweenness\cite{Korniss,Newman,Brandes,Ravasz_2010}:

\begin{equation}
\ell_{ij} = \frac{1}{N-1} \sum_{s,t=1}^{N} |I_{ij}^{(st)}|,   \ell_{i} = \frac{1}{N-1} \sum_{s,t=1}^{N} |I_{i}^{(st)}|.
\end{equation}

In order to obtain the  $I_{ij}^{st}$ currents along the edges from one source/target pair, we use Kirchhoff's law for each node $i$ in the network and solve the system of linear equations:

\begin{equation}
\sum_{j=1}^{N} A_{ij}\left( V_{i} - V_{j} \right) = I \left( \delta_{is} - \delta_{it} \right),   \forall i = 1,...,N.
\label{lapl}
\end{equation}

Here, we assume that $I$ units of current flow through the network from source $s$ to target $t$, and $A_{ij}$ denotes the adjacency matrix of the network. This equation can be rewritten in terms of the weighted network Laplacian $\mathcal{L} =\delta_{ij}k_{i}-A_{ij}$, where $k_{i}=\sum_{j}A_{ij}$ is the degree of node $i$. Thus, we can write Eq.~\ref{lapl} as $\mathcal{L}V=\mathcal{I}$ , where $V$ is the unknown column voltage vector, and $\mathcal{I}_{i}$ is the net current flowing into the network at node $i$, and takes nonzero values only for the source and target nodes. Since the $\mathcal{L}$ network Laplacian is singular, we find the pseudo-inverse Laplacian $G=\mathcal{L}^{-1}$ using spectral decomposition \cite{Korniss_2006,Korniss,Hernandes}. Thus, by choosing as reference potential the mean voltage \cite{Korniss_2006}, $\hat{V}_{i}=V_{i}-\langle V \rangle$, where $\langle V \rangle=(1/N)\sum_{j=1}^{N}V_{j}$ for each node $i$ we obtain:

\begin{equation}
\hat{V} = \left( GI \right)_{i}=\sum_{j=1}^{N} G_{ij}I \left( \delta_{js} - \delta_{jt} \right)=I\left( G_{is} - G_{it}\right)
\end{equation}

Therefore, for $I$ units of current and for a given source/target pair, the current flowing through edge $(i,j)$ can be written as

\begin{equation}
I_{ij}^{(st)} = A_{ij}\left( V_{i} - V_{j} \right) = A_{ij} I \left( G_{is} - G_{it} - G_{js} + G_{jt}\right).
\end{equation}

The above equation shows that current along an arbitrary edge is uniquely determined by network topology.

In modeling of the the electrical flows in the power grid a commonly used approach is the use of the DC power flow model \cite{Bernstein_2011,Verma,Scala,Rahnamay,Zimmerman}, where links, in addition to resistance, also possess reactance. However, it has been shown in \cite{Bernstein_2011} that the equations for this DC model of power flow bear a close resemblance to that of an analogous electrical circuit. In prior studies it has also been demonstrated that, despite neglecting the true AC nature of the power grid, inferences made by employing the DC power flow model can still be useful\cite{Scala}.

It is important to point out that our goal is to study the fundamental aspects of cascades on spatial networks carrying ditributed flow, not designing strategies specifically tailored for electrical power transmission systems.

\subsection*{Cascade model}
We simulate the cascading failures in our networks using the model of Motter and Lai \cite{Motter_2002}. We assign each node a load-bearing capacity proportional to its initial state load in the form of $C_{i}=(1+\alpha)\ell_{i}^{0}$, where $C_i$ is the capacity of node $i$, $\ell_{i}^{0}$ is its initial state load when the network is intact, and $\alpha$ is a tolerance parameter. When a node fails, its load is redistributed among the surviving nodes. Nodes that after this redistribution are assigned a higher load than their capacity $\ell_{i}>C_{i}$ also fail, resulting in further redistribution of load, and possible overload of other nodes. Such an avalanche of subsequent failures is called a cascading failure.

\subsection*{Empirical network}
Our work is centered on the study of the UCTE European power transmission network from year 2002
\cite{Zhou_2005,Zhou_2013,UCTE_data}. The network comprises 18 European countries, contains $N=1254$ nodes (buses) and $E=1812$ edges (transmission lines). The average degree of the network is $\langle k\rangle=2.89$.

\subsection*{Network models}
In addition to our study on the empirical UCTE power transmission network, we also test our methods on various types of artificial networks. Below, we briefly describe our approach in generating these synthetic network ensembles.

\textbf{Random geometric graphs (RGG)}
We construct RGG graphs of size $N$ in 2D by placing $N$ nodes randomly in the unit square with open boundary conditions, and connecting any pair of nodes if the Euclidean distance between them is less than the connection radius $R$ \cite{Penrose_2003,Dall_2002}. Based on $\langle k \rangle=\pi R^{2}N$ formula, we control the average degree $\langle k \rangle$ of the graph by varying $R$.

\textbf{Scale-free (SF) networks}
We construct scale-free networks \cite{Barabasi} of size $N$ by first generating a degree sequence from a prescribed power-law degree distribution $P(k)\sim k^{-\gamma}$ that yields a desired average degree $\langle k \rangle$. Next, we pass this degree sequence to the configuration model \cite{Molloy} in order to obtain the scale-free network. Here, we consider the degree sequence as a list of the number of edge stubs (half-links) each node has, and in the configuration model we randomly choose two nodes with available edge stubs and connects them to form an edge. The process is repeated until all edge stubs are connected.

\textbf{Erd\H{o}s-R\'{e}nyi (ER) graphs}
We construct ER graphs \cite{Bollobas} of size $N$ by connecting every pair of nodes with probability $p$, parameter used to control the average degree of the network using $\langle k \rangle=p(N-1)$ formula.

\subsection*{Power-law statistical testing}
For statistical testing of the power-law fits, we use Kolmogorov-Smirnoff (KS) test \cite{Clauset_2009} for goodness of fit. As the authors point out, the least-squares method can give inaccurate estimates of parameters for power-law distribution, and in addition, they give no indication of whether the data does indeed follow a power law \cite{Buldyrev_pl}. Therefore, based on this seminal work, we employ statistical testing using the maximum-likelihood fitting (MLE) to estimate the accuracy of the fitted power-law, and KS test for validating the goodness of fitting a power-law. Our empirical results are obtained as probability density function (PDF), denoted as $p(X)$ (see Suppl. Info. S9), however the KS-statistic requires the cumulative distribution function (CDF), which we obtain as $P_{>}(S)=\sum_{X=S_{min}}^{S}p(X)$.
Finally, we extract the slope by performing a least-squares linear regression on the logarithm of the cumulative distribution function. For a more detailed description, see Suppl. Info. S9.

\subsection*{Outlier removal}
We calculate the prediction bands and confidence intervals using statistical methods (see Suppl. Info. S10), and mark as outlier every data point that falls outside the prediction bands. After we remove them, we fit a linear slope to analyze the correlation between the two variables.

%

\section*{Acknowledgements}

The authors would like to thank Andrea Asztalos and Sameet Sreenivasan for valuable discussions. This work was supported by DTRA Award No. HDTRA1-09-1-0049, by the Army Research Laboratory under Cooperative Agreement Number W911NF-09-2-0053 and by NSF Grant No. DMR-1246958. The funders had no role in study design, data collection and analysis, decision to publish, or preparation of the manuscript.

\section*{Author Contributions}
A.M., N.D., X.L., B.K.S. and G.K. designed the research;
A.M., N.D., X.L. implemented and performed numerical experiments and simulations;
A.M., N.D., X.L., B.K.S. and G.K. analyzed data and discussed results;
A.M., N.D., X.L., B.K.S. and G.K. wrote and reviewed the manuscript.

\section*{Additional Information}
Competing financial interests: The authors declare no competing financial interests.

\newpage

\section*{Figures \& Captions}

\pagebreak

\end{document}